\documentstyle[12pt]{article}
\newcommand{\hi}{{H \sc i}\ }
\newcommand{\hii}{{H \sc ii}\ }
\newcommand{\kms}{\rm km s$^{-1}$}
\newcommand{\etal}{\rm et al.\ }

\begin{document}
\vspace{4cm}
\large
\centerline {\bf DISPERSAL AND MIXING OF OXYGEN}

\centerline {\bf IN THE INTERSTELLAR MEDIUM}

\centerline {\bf OF GAS-RICH GALAXIES}

\bigskip
\bigskip
\normalsize
\centerline {\bf Jean-Ren\'e Roy${\rm ^{1,2)}}$ and Daniel Kunth${\rm ^{3)}}$}
\vspace {1cm}
\noindent
\normalsize
1) D\'epartement de physique and Observatoire du Mont M\'eganic,
Universit\'e Laval, Qu\'ebec, QUE G1K 7P4, Canada \\
\noindent
2) Observatoire de Paris, DAEC, Unit\'e associ\'ee au CNRS, DO 173,
et \`a l'Universit\'e Paris 7, 92195 Meudon Cedex, France\\
\noindent
3) Institut d'Astrophysique, 98 bis, Blvd. Arago, 75014 Paris, France \\

\vspace{4.5cm}
\noindent
\centerline {Astronomy \& Astrophysics -- Main Journal} \\

Received 17 March 1994/ Accepted
\vspace{3cm}
\vfill
\noindent {\tt Send all correspondence to :}
\noindent Jean-Ren\'e Roy, D\'epartement de physique, Universit\'e Laval,
Qu\'ebec, QUE G1K 7P4, Canada ({\tt jrroy@phy.ulaval.ca})
\newpage
\centerline {\bf Abstract}

\bigskip

\bigskip

Stellar and nebular abundance indicators reveal that
there exists significant abundance fluctuations in the interstellar
medium (ISM) of gas-rich galaxies.  It is shown that at the present
observed solar level of O/H $\sim 6 \times 10^{-4}$, abundance
differences of a factor of two, such as existing between the Sun
and the nearby Orion Nebula, are many times larger than expected.
  We examine a variety of
hydrodynamical processes operating at scales ranging from  1 pc
to greater than 10 kpc, and show that the ISM should appear better
homogenized chemically than it actually is: $(i)$ on
large galactic scales (1 $\geq\ l\ \geq$ 10  kpc), turbulent diffusion of
interstellar clouds in the shear flow of galactic differential rotation
is able to wipe out
azimuthal O/H fluctuations in less than $10^9$ yrs;
$(ii)$ at the intermediate scale (100 $\geq\ l\ \geq$ 1000 pc),
cloud collisions and
expanding supershells driven by evolving associations of massive
stars, differential rotation and triggered star formation will
re-distribute and mix gas efficiently in about $10^8$
yrs; $(iii)$ at small scales (1 $\geq\ l\ \geq$ 100 pc),
turbulent diffusion may be the dominant mechanism in cold clouds, while
Rayleigh-Taylor and Kelvin-Helmhotz instabilities quickly develop in
regions of gas ionized by massive stars, leading to full mixing in
$\leq 2 \times 10^6$ yrs.

It is suggested that the relatively large O/H fluctuations observed  in
large disk galaxies may be due to retention, in sites favored by triggered star
formation,  of freshly enriched ejecta from
SNR and supershells expanding in a differentially rotating disk,
plus, possibly, $infall$
of low metallicity material from individual clouds
like high velocity clouds which splash on the disk on
timecales shorter than the local mixing time. In low-mass galaxies,
stimulated star formation is much less efficient, and the most
effective mixing mechanisms are absent; the escape  of newly
enriched material due to  galactic winds powered by the starburst events,
the lack of large-scale stirring,
and the long dormant phase between successive star forming episodes
make possible the survival of large abundance discontinuities.

\bigskip


\vfill
\newpage

\noindent
{\bf 1. Introduction }\\

\bigskip

If one excludes
the global radial abundance gradients observed in galactic disks,
the ISM of the more massive galaxies appears on first examination relatively
 well-mixed.  The chemical composition
of the interstellar medium (ISM) in large spirals at a given radial distance
appears  not to vary much, but there are
evidences that the ISM is not perfectly homogenized. O/H abundances
derived from several indicators in large disk galaxies suggest that azimuthal
variations could reach
a factor of two (Belley \& Roy 1992). However this is hard to quantify further
 because of the uncertainties underlying the
procedures of nebular abundance derivations.
Low-mass galaxies have
very shallow global abundance gradients,
and variations of O/H from one region
to another in magellanic irregulars are moderate with fluctuations
by a factor of about two (Dufour 1986), indicating
that mixing of the ejecta from massive stars is not perfect in
these systems either.

It is true that
one does not see, in relatively massive galaxies, isolated pockets of ionized
gas with O/H = 1/50
O/H$_\odot$ close to regions with O/H = 2 O/H$_\odot$; thus mixing processes
appear
quite effective. How effective they are is a relevant  question since the sites
of oxygen production
could  be correlated by triggered star formation, and give rise
to significant spatial abundance fluctuations. The largest
 molecular clouds may survive the destructive photodissociation effect of the
massive stars
to which they give birth, and self-enrichment or auto-pollution of these clouds
could occur (Gilmore 1989); however the survival of molecular clouds to
evolving massive stars or to passage through spiral arms

remains questionable. On the other hand, the mixing in low-mass galaxies
might be less efficient.  This is strongly suggested by
the largest known spatial
discontinuity in heavy element abundances measured recently
in the dwarf galaxy I Zw 18 by Kunth \etal (1994); employing
high-resolution spectroscopy on GHRS-HST, they have derived the oxygen
abundance in the cold cloud associated
with the main star forming region of this well-kown dwarf
galaxy.  While the ionized
gas has O/H $\approx$ 1/30 O/H$_\odot$, the measured
O/H in the associated cold cloud is 30 times less,
$i.e.$ 1/1000 solar!

The importance of inhomogeneous
chemical evolution of galactic disks has been emphasized
recently by Malinie \etal (1993) who  showed that chemical inhomogeneities
provide a much better fit to the abundance distribution function
of G-dwarfs in the solar neighborhood. Poor mixing of the interstellar
gas has also
been proposed by Lennon \etal (1990) as an explanation for the
observed discrepancies between the large-scale abundance gradients
in our Galaxy deduced from \hii regions (Shaver \etal 1983), and
from hot stars (Gehren \etal 1985; Fitzsimmons \etal 1990);  indeed B-type
stars
observed between about 5 and 15 kpc do $not$ show systematic abundance
variations.

In this paper, we wish to address the issue of dispersal and mixing
of newly-formed elements in the interstellar medium by reviewing
 the various mechanisms
which are responsible for the
 chemical homogenization of the interstellar gas,
or the suppression of it both on local and galactic scales.
This problem has been recently examined by Bateman \& Larson
(1993). They show that cloud motions may be the dominant
mechanism for the dispersal of Fe.
We will show that significant spatial abundance fluctuations exist
in galaxy disks despite the apparent efficiency of mixing,
and that these fluctuations are likely to be largest
in very  low-mass galaxies.

\bigskip
\noindent {\bf 2. Evidence for abundance variations}\\

\bigskip

Pagel (1993) has reviewed the abundances of heavy elements in the
solar neighborhood, for
which solar abundances are believed to be accurate to within $\pm$ 0.1 dex.
The range of oxygen abundances is 0.3 dex as derived from ten
different Galactic standards; how much of this is due to errors in the
methods is not clear.
For the ISM of the Galaxy, the  abundance surveys of
Shaver \etal (1983) and  Fich \& Sulkey (1991) are consistent with O/H
variation up to a factor of 2 over a scale of about 1 kpc, best illustrated
 by the Sun-Orion Nebula well-known difference for
the abundance of oxygen; the detailed study by Baldwin \etal
(1991) gives O/H in the Orion Nebula as being 0.44 solar. The precision
of abundance measurements in other galaxies is often hampered by the 0.2 dex
uncertainty of the
semi-empirical method of using the ratio of bright nebular lines. Nevertheless
 the dispersion at a given galactic radius of derived O/H values
appears much larger than the experimental uncertainties. For the well-sampled
galaxy NGC 2997, for example,
azimuthal variations by a factor of two cannot be excluded
$(cf.$ Walsh \& Roy 1989).  Detailed studies of the chemical
compositions of \hii regions in the Magellanic Clouds are also consistent
with overall abundance fluctuations by a factor of two (Dufour \& Harlow 1977;
Pagel \etal 1978).

Abundances have been measured in a large number of stars with a high degree
of accuraccy; these measurements reveal real scatter, $i.e.$ which
is larger than the observational errors and uncertainties $(cf.)$ Edvardsson
\etal 1993). The dispersion in the age-metallicity relation of nearby F stars
(Carlberg \etal 1985), at least
for the youngest stars, is indicative of interstellar abundance
inhomogeneities.
The scatter in stellar metallicities is very likely
reflecting the original inhomegeneities of the interstellar gas
(Gilmore 1989; Fran\c cois \& Matteucci 1993).
The existence of real scatter in the enrichment of the interstellar medium
at any time is also seen in the age-metallicity relation
for nearby clusters and stellar groups observed by Boesgaard (1989) who
concluded that  the apparent lack of an age-metallicity relationship
indicates that  the enrichment and mixing in the Galactic disk have not been
uniform on timescales less than 10$^9$ yr.  Rolleston \etal (1994) have derived
stellar abundances of B stars belonging to clusters separated by distances
of the order of 1 kpc and found abundance variations of a factor of five. From
their extensive
analysis of 189 nearby field F and G disk dwarfs, Edvardsson \etal (1993)
find a residual scatter of probably 0.15 dex in [Fe/H] at a given
radius which remains
to be explained by mechanisms other than simple chemical evolution.
They also find that the $\alpha$-group elements follow the iron group
elements very closely; there is no significant scatter in [$\alpha$/Fe] at a
given
age and galactocentric distance, which indicates that the nucleosynthetic
products of the supernovae of different types are mixed
locally in the interstellar gas.

Fluctuations by a factor of two correspond to $\delta$O/H $\sim$ 3 $\times$
10$^{-4}$. How do these compare to what would be expected
in a perfectly mixed disk? This calculation was done some time ago by
Edmunds (1975), and it is interesting to repeat here his reasoning,
using updated values for some of the physical parameters.
We first assume that massive stars are randomly distributed in space
and time in order to derive what would be the expected abundance
fluctuations in a fully mixed ISM enriched only by $isolated$
stars. We suppose then that every massive star
develops a wind-blown bubble into which it finally explodes as a supernova.
The winds and explosions
act as an effective mixing process (as we will see later) as well
as providing new metals. From the
$\Sigma - D$ diagram of Galactic supernova remnants, we estimate
that each remnant reaches a radius of about 50 pc before it becomes
undistinguishable from random structures in the disk (see Green 1984).
 Let us assume a simple homogeneous
model of the Galaxy as a disk of  R = 12 kpc and z = 200 pc; each wind bubble
\&
supernova remnant will occupy a fractional volume $\sim 6 \times 10^{-6}$
of the disk. Assuming a constant SN rate of 1/100 yr$^{-1}$ over $\sim10^{10}$
yrs, then at any point in the disk, an average of at least 580 SN events would
have contributed by now to the metal enrichement of the ISM. We suppose that
the metals ejected in the local ISM are fully mixed within
the 50 pc radius during the lifetime of the remnant,  and that
the production of SN is randomly distributed throughout the disk.
If $n$ events contribute to give aproximately the solar O/H abundance of
6.9 $\times$ 10$^{-4}$, then fluctuations of $\delta$O/H
$\sim$ (O/H$_\odot$)/$\sqrt n \sim\ 3 \times 10^{-5}$ would be expected.
This is at least ten times smaller than observed, and considering our
simplistic assumptions, the
result is not surprising.

The above scenario is unlikely to be appropriate, because
massive star do not occur randomly; instead their formation is highly
correlated in space and in time (see the review by
Tenorio-Tagle \& Bodenheimer 1988). This reduces the number of required
enrichment events, and leads to an increase in the expected fluctuations.
A visual inspection of
any large gas-rich disk galaxy shows that, at any given time, of the
order of $\sim10^3$ massive star formation sites are active, and that
they occupy $\sim 10\%$ of the volume of the disk. We assume that the star
formation
rate over the last 10$^{10}$ years has been on average equal to the present one
(Kennicutt 1994: private communication);
the  lifetime of each starburst is $\sim 2 \times 10^7$ yrs.
Thus about n =  50 enrichment events would have taken place in any given
volume of the disk. Although this number is much lower than
in the random scenario, it does not
explain the observed large O/H spatial fluctuations such as the Sun-Orion
difference. Furthermore, this stationary scenario is also extreme, because
it excludes the homogenizing effect of any mechanism capable of mixing
the ejecta on scales larger than the individual \hii regions.
Consequently this discussion suggests that the observed fluctuations are
many times larger than expected.

\bigskip
\noindent {\bf 3. The mixing mechanisms for the gas}\\

\bigskip
In this section, we review the various mechanisms responsible
for mixing the ISM on various scales, and we show that there is a rich variety
of processes capable of mixing the ISM fully and efficiently; these
mechanisms are especially effective in the
ionized gas of star forming regions. We investigate these mechanisms on
three scales: large (1 - 10 kpc), intermediate (100 - 1000 pc) and small
(0.1 - 100 pc).

\bigskip
\noindent 3.1.  Large-scale transport, differential rotation and azimuthal
homogenizing

\bigskip
Like stars, gas clouds follow complex orbit patterns around
the galaxy center. But as opposed to stars, clouds interact
with each others; \hi clouds are scattered isotropically
when colliding with each other every $10^7$ yrs (Spitzer 1978, Hausman 1981),
and massive molecular clouds travel
"straight" before being blown apart by massive stars. Roberts \& Hausman
(1984) have estimated the mean
free path for molecular clouds to be in the range $l$ = 300 -- 1000 pc.

In a galaxy with differential rotation, motions
around the center can be approximated by the superposition
of a retrograde motion at angular frequency $\kappa$ around
a small ellipse with axis ratio 1/2, and prograde motion of the
ellipse's center at angular frequence $\Omega$ around
a circle (Binney \& Tremaine 1987). For example, at the
solar Galactic distance from the center, stars make about
1.3 oscillations in the radial direction to complete an orbit
around the center in the time they need to complete their orbit, $i.e.$
the epicyclic period is $\sim 1.8 \times 10^8$ yrs and
 stars make typical radial excursions of about 800 pc (Tayler 1993).
Their radial excursions will appear, to an observer in circular
orbit, as drifts at velocity $v$.

While stars keep their identity, clouds collide, merge or disperse.
At a given point in the disk, one can imagine the various clouds as belonging
to various orbits, criss-crossing the reference circular orbit.  Because
intercloud collision times take place  on time scales
much shorter than the orbital period,
clouds loose their identity; individual parcels
of gas will  effectively change orbits and mix, while carried by the large
scale
flow around the galaxy. The net effect of cloud-cloud collisions is to act as a
scattering
process allowing clouds, fragments of clouds or parcels of gas to jump to
different orbits describing each their own rosette; their overall motions
correspond
to nonstationary turbulent transport in a shear flow, the shear
being caused by the differential rotation of the galaxy.
In such a case, diffusion is much more effective in the direction of the mean
flow -- direction of rotation -- than in the directions perpendicular to it.

Turbulent transport in shear flows is well discussed by Tennekes and
Lumley (1983). Consider a cartesian coordinate system
centered at a given galactocentric point at distance R from the galaxy
center. Let us take $x_1$ to be the direction tangential to the local
circular orbit,
and $x_2$ be the radial direction; we neglect effects perpendicular
to the galactic plane. Let us suppose that the shear
is due to differential rotation, and that orbiting clouds
crossing the local circular orbit at a given point
appear to come from random directions at rms velocity $v$;
$v$ includes also the velocity dispersion of
interstellar clouds.
$S$ is the radial gradient in velocity due to
differential rotation; $l$ is the  mean free path of clouds. In the direction
perpendicular to the orbit, the
time to diffuse a length scale $\Delta x_2$ is given by stationnary turbulence
as

\bigskip

\centerline {$ \tau_{x2} \ = \ {\Delta x_2^2  \over v \ l}. $}

\bigskip

\noindent But as a wandering parcel of gas moves in the $x_2$ radial direction,
it moves into a region with a different mean velocity, so that it tends
to move faster (or slower) than in a flow without shear. Consequently,
in the orbital direction $x_1$, the time to diffuse downstream and
upstream is given (Tennekes \& Lumley 1983) by

\bigskip

\centerline {$\tau_{x1} \ = \ {\Delta x_1^{2/3} \over S^{2/3} \ v^{1/3}
\ l^{1/3}}$.}

\noindent Thus the dispersion in the downstream and upstream direction
(the direction of rotation) increases much faster than
the dispersion perpendicular to it. This is the key to efficient
azimuthal mixing in large disk galaxies with strong rotational velocity
field.

Suppose star associations A and B being on the same orbit
but diametrically opposed; at a given time, they suddenly enrich their
neighborhood
by exploding into supernovae. Using the last equation, we can calculate the
time for collisions of clouds and differential rotation to diffuse their
patches of enriched elements into each other
($i.e.$ after travelling each 1/4 of a full orbit length). We suppose $S$ = 10
\kms \ kpc$^{-1}$, $l$ = 300 pc, and $R_G$ = 10 kpc; we choose the value of $v$
taking
a typical star velocity with respect to the local standard of
rest; the latter defines the rotational speed of a hypothetical set of
stars in precisely circular orbits. Taking  $v \approx 10$ \kms
(Mihalas \& Binney 1981), the timescale for mixing clouds azimuthally in a
galaxy is $\leq$10$^9$ years. Consequently,
galactic clouds orbiting at roughly the same
galactocentric distance will fully mix in a
small fraction of a Hubble time; this ``epicyclic'' mixing  or dynamic
diffusion is able to erase any azimuthal abundance variations in large disk
galaxies.

\bigskip
\noindent 3.1.1  Bar-induced mixing or the role of radial flows

Bars can induce strong radial flow in the interstellar gas
of a galaxy (Lacey \& Fall 1985, Struck-Marcell 1991, and
Friedli \& Benz (1993).  Vila-Costas \& Edmunds (1992), and
Zaritsky, Kennicutt \& Huchra (1994) have shown that
global O/H gradients in intermediate and barred galaxies
are shallower than gradients in normal galaxies. Martin \& Roy
(1994) have demonstrated that abundance gradients become flatter
as the length or the ellipticity of the bar increases,
$i.e.$ the stronger the bar is, the flatter the abundance
gradient becomes. The most direct explanation for this relation is that
the strong radial flow associated with the bar acts
as an efficient homogenizer of the chemical
composition in the interstellar medium. Indeed
numerical simulations by Friedli \& Benz (1993) demonstrate
that radial flows of several tens \kms \ can operate
over a large fraction of the galaxy disk.

Magellanic systems display in general a strong bar and have flat
abundance gradients (Edmunds \& Roy 1993). We suggest
that in such systems, differential rotation provides some
degree of azimuthal homogenization, while the action of a bar would
homogenize the gas radially
on a timescale of less than 1 Gyr (Friedli, Benz \& Kennicutt 1994).

\bigskip

\bigskip
\noindent 3.2. Intermediate-scale transport

\bigskip
At the 100 - 1000 pc scale, intercloud collisions will be efficient
contributors to mixing, since their mean free path is of the same
order of scale. In addition, propagating star formation
triggered by expanding shells driven by evolving massive stars
could transport and mix interstellar gas  $radially$ over
a galaxy disk.  We discuss this scenario
in more details because it may be the dominant diffusion process
in galaxies with weak rotational field and continuous star formation.

McCray \& Kafatos (1987) and Elmegreen (1992) review
several cases of possible supershell-induced star formation.
The close association of \hii complexes with large H I holes observed in the
Large
Magellanic Cloud (Meaburn \etal 1991; Dopita \etal 1985),
in Homberg II (Puche $etal.$ 1992), IC 2574 (Martimbeau
\etal 1994) and in NGC 6946 (Boulanger \& Viallefond 1993)
suggests that triggered star formation is taking place
in expanding  supershells.
The net effect of this activity is the shuffling and re-distributing of the
interstellar gas over relatively large scales.

In massive disk galaxies, Edmunds (1975) and Palou\u s \etal (1990)
have shown that galactic rotation
will shear the mixing volume of expanding interstellar bubbles into
elliptical shape stretched in the direction of rotation about the
galactic center, due to the streaming effect discussed in section 3.1.
Typical amount of stretching  are illustrated
by the semi-minor and semi-major axies of the largest HI holes in M31 and
M33 listed in Palou\u s \etal which are 190 $\pm$ 80 pc and
410 $\pm$ 300 pc over a typical time of $\sim$ 6.7 $\times$ 10$^7$ yrs.
In other words, interstellar gas caught in an expanding
supershell in a low mass galaxy will be displaced into an elliptically
stretched expansion as shown by the
simulations of Palou\u s \etal (1990).
A question, discussed in section 4.1, is whether the new nucleosynthetic
products polluting the expanding shells will fully disperse and mix on
 a larger scale $before$ the parcels of gas of
 expanding superbubbles enter a new star formation episode again.

To quantify the transport of gas parcels through successive star forming
episodes, we
treat this as a diffusion problem in the case of homogeneous,
stationary turbulence.
We assume that most of the gas stays in the disk at all times, that is
we exlude "breakout" or "blowout". Bursting out of the galaxy disk may happen
in the cases of very large
star formation events, where most of the
momentum and kinetic energy escapes as a galactic wind, preventing runaway
transformation of all the remaining gas into successive
generations of stars.
"Breakout" or "blowout" of superbubbles out of the
galaxy disk, result in internal pressure decreasing  so much
that the accumulation of gas and formation  of new clouds in the galaxy
plane may stop. This is more likely to happen in the very low mass galaxies
where escape velocities are only $\sim 100$ \kms .

The collisional
mean free path, $\lambda$, is the average distance traveled by
a parcel of gas before being caught in another star forming
event which will drive its own supershell; it is assumed
to correspond to the typical radius of \hi \ holes which is
$\approx$ 250 pc in  small galaxies like
Holmberg II (Puche \etal 1992), and 100 pc in large spirals
like M 31 (Brinks \& Bajaja 1986) or M 33 (Deul \&
Den Hartog 1990; Court\`es \etal 1987). These values
are larger than 50 pc used in section 2 for shells associated with single
stars;
this is because supershells are the results of evolving OB associations
which contain many massive stars.
The velocity, V, of the parcel
of gas which is being propelled by a large  expanding bubble
is of the order of 40-50 km/s (Roy $etal.$
1991, 1992) during the early phase. This velocity drops to 5-10 km/s
in the latter evolutionary stages of the
expanding bubble (Puche $etal.$ 1992).
We use the higher value of velocity and we consider that the
parcel of gas lingers at low velocity or remains quasi-stationnary for
a certain period of time that we will discuss below. If moving continuously
all the time,
a  parcel of gas would random walk across the radius ($R_G$) of the galaxy
 in a time given by

\bigskip
\centerline {$\tau_d~\approx~ {{R_G}^2\over V~\lambda}$.}

\bigskip\noindent
If $R_G$ = 4 kpc ($e.g.$ magellanic irregulars), V = 45 \kms, then
$\tau \sim 1.5 \times 10^9$ years for a small galaxy ($\lambda \sim$ 250 pc),
and
$\sim 10^{10}$ years in a large galaxy ($\lambda \sim$ 100 pc; $R_G$ 10 kpc).

\bigskip

This value of $\tau_d$ is
a lower limit because a crucial aspect is missing. Triggered
or supershell-induced star formation is not a continuous process,
$i.e.$ the parcel of gas
may find itself in a new region of the interstellar medium where
gravitational collapse, thus  star
formation, is not immediate. Some
time may elapse before the onset of the next self-gravitation instability;
this is the dormant phase.
We need first to find what fraction of
time a gas parcel spends in a starbursting region compared to
the time spent in the dormant phase of stable \hi filaments.
 From the above numbers,
one could expect radial homogenization in less than a Hubble time
in small galaxies, only if a parcel of gas spends more than  20\% of
its time in expanding supershells. The length of time spent by the parcel
of gas in expanding supershells, $i.e.$ the time spent moving around
can be estimated from published numerical simulations.
Igumentshchev $et al.$ (1990) among others have modeled the evolution of large
expanding shells generated by the collective winds
and sequential supernovae from OB associations.
Their calculated supershells, with radius and velocity corresponding
to observed ones, have lifetimes $t_{shell}$ =  13 to 16 Myr; these are
shorter than the apparent ages of the largest \hi holes.

On the other hand, the dormant phase includes the duration of gravitationnal
collapse and the period where the gas remains at rest; the later is very
difficult to evaluate. We first discuss the timescale for collapse.
Elmegreen (1992, 1994) has investigated triggered star formation along the
perimeters of expanding giant shells by analyzing the radial and transverse
gravitational collapse of such structures.
The collapse of swept-up matter in an
expanding and decelerating shell give rise to instabilities obeying
a dispersion relation which allows to derive the time and the radius for
collapse of clouds in a shell
expanding with velocity V in a medium of a given pre-shell density;
the time and distance for significant collapse are  given
by Elmegreen (1992, 1994) as

\bigskip
\centerline {$t_c\ = \ 103 \bigl({n_0 \ M \over{\rm cm}^{-3}}\bigr)^{-1/2}
\ \ {\rm Myr,}$}

\bigskip\noindent and,

\centerline {$ R_c\ = \ 176 \ M^{1/2}\ \bigl({c \over{\rm km/s}}\bigr)
   \ \bigl({n_0 \over{\rm cm}^{-3}}\bigr)^{-1/2} \ \ {\rm pc,}$}

\bigskip\noindent
where $n_0$ is the preshell number density, $V$ is the
instantaneous expansion speed, $c$ is the rms velocity
dispersion in the shell; for normal disk conditions,
$M$ = V/c $\approx$ 2. (For the
collapse of an expanding ring, the equations are similar
and will be discussed in section 4.1).
Assuming an average pre-shell number density of $\sim$ 10 cm$^{-3}$
(from N(HI)/$d$ $\approx$ 10$^{21}$ cm$^{-2}$/400 pc)), c = 10 km/s,
we find $t_c$ = 20 Myr and $R_c$ = 750 pc; this size is consistent
with the radii of the largest \hii complexes observed
which are $\sim$ 0.5  kpc. We assume that the
$minimum$ duration of the dormant
phase is given by $t_c$; this time is of the same order as the duration of
the expanding phase.

Therefore the crucial timescale is the time spent by the gas doing nothing.
There is no direct way to estimate this, except to consider that
large disk galaxies and magellanic irregulars have a constant SFR;
thus the fractional area occupied by present
day \hii regions leads to an estimate of the fraction of time the
gas spends in the dormant phase. Since \hii regions cover about
10\% of disk galaxies, the gas spend about 90\% of its time not moving.
Thus full radial mixing may take place in a time of the order
of the Hubble time only in the smaller galaxies (magellanic irregulars
and less massive system with $continuous$ star formation).
Nevertheless superbubble expansion will contribute to mix gas on scales
of $l \leq 1$ kpc in a timescale of 10$^8$ years in all galaxies
where massive star formation is taking place.

\bigskip
\noindent
3.3. Small-scale mixing

\bigskip
\noindent
3.3.1  Turbulent diffusion

\bigskip
Turbulent diffusion takes place in ionized, neutral and
molecular clouds. Turbulent diffusion is to be opposed to
molecular diffusion, which is much slower and less efficient
than eddy diffusivity acting when turbulent motions are present.
In a medium with characteristic length scale L,  $i.e.$
motions are present of scales $\leq$ L, and with a characteristic
velocity $u$ (defined as the rms velocity fluctuation in the
medium), the time scale for diffusion is $\tau_{\rm diff}\ \sim \
L/u$. Taking $u$ = 20 \kms \ (Arsenault \& Roy 1988)
for the ionized gas, $u$ = 2 \kms \ (Kulkarni \& Heiles 1988)
for the neutral gas clouds and $u$ = 0.5 \kms \
(Falgarone \& Phillips 1991) for the
cold molecular clouds, one derives characteristic times
for simple diffusion of 4 $\times\ 10^7$ yrs, 4 $\times\ 10^8$
yrs and 1.6 $\times\ 10^8$ yrs for ionized, neutral and molecular clouds
respectively;
here we have assumed $L \ \sim$ 1 kpc for ionized and neutral clouds, and
$ L \ \sim$ 100 pc pour molecular clouds.
For neutral and molecular gas, these timescales are much
longer than the lifetimes  (10$^7$ years) of the clouds themselves; thus
inhomogeneities will not have the time to smooth out, before
the destruction of the clouds by collision or star forming events.
Thus homogenization must occur during the HII region lifetime to ensure
 mixing of individual clouds. It is likely that large abundances
anomalies could survive for up to 10$^9$ years in the cold gas.

\bigskip

\noindent 3.3.2 Fluid instabilities as mixing mechanisms

\bigskip

For \hii regions, the development
of Rayleigh-Taylor (R-T) and Kelvin-Helmholtz (K-H) instabilities
takes place over a relatively short timescale and
permits thorough mixing over the lifetime of star forming regions.

First several energetic phenomena associated with star formation events
may give rise to instabilities in gas flows; they are related to
mass losses from forming or evolving stars, and
from moving stellar or nebular enveloppes.
The development of these instabilities  makes the
surrounding ISM fully turbulent, if the growth time of the
instabilities is shorter than the lifetime  of the object.
Relevant phenomena are collimated outflows from young stars
(Schwartz 1983), explosive ejection of matter associated
with star formation (Allen \& Burton 1993), champagne
flows from \hii regions bursting out of their parent molecular
clouds (Tenorio-Tagle 1979), expanding supershells due to supernovae and
stellar winds from star clusters, and superbubbles
from evolving OB associations.  The interaction of these
dynamical structures with the ambient interstellar medium
can give rise to the Rayleigh-Taylor and Kelvin-Helmholtz
instabilities which develop over the surface of the
moving components. The Rayleigh-Taylor instability
leads to the fragmentation of the moving components,
$i.e.$ expanding shells break apart, and pieces
of increasing size break away from the front of
a collimated jet.  The Kelvin-Helmholtz instability
develops along the sides of the flow at the sheared
surface of two fluids moving differentially with
respect to each other.  Once the waves have grown
sufficiently, there is a shearing of the wave leading to vorticity,
$i.e.$ to a growing boundary of turbulent eddies.

The growth times of the surface waves, measured in terms
of their e-folding times, can be calculated for an expanding shell driven
by winds and supernovae. Chandrasekhar (1961; Shore 1992) give
relations which can be used to calculate the growth
time and the growth speed of the surface waves in the absence
of a magnetic field;  the relations for an imcompressible
fluid are applicable for the cases in hand.
If a denser material with $\rho_2$ is accelerated into a less dense one with
$\rho_1$, some material from the two layers will be exchanged when
a perturbation occurs at the interface; this will automatically happen
when the density gradient is the opposite direction to the local
acceleration. The upward and downward displacements of fluids will produce
a change of potential energy equal to the total kinetic energy gained, $i.e.$

\centerline {$ (\rho_2 - \rho_1)\ g \ \delta z \ =\ (\rho_2 + \rho_1)\ \delta
z^2
\ \tau_{R-T} ^{-2}$}

\noindent where $\tau_{R-T}$ is the growth rate of the disturbance and $\delta
z$ the displacement.
Thus the growth time of Rayleigh-Taylor instabilities is
simply given by

\bigskip
\centerline {$ \tau_{R-T} \ = ({\rho_1 + \rho_2\over \rho_2 - \rho_1})\
(gk)^{-1/2}$}

\bigskip\noindent
where $g$ is the effective gravity and $k$
is the wavenumber of the largest instability. Let us take the case of a
large expanding shell pushed outward by evolving massive stars.
For a continuous energy input, $ R \propto t^{0.6}$
(Castor \etal 1975), and $g = {2 \over 3} {v^2\over R}$.
Taking 100 pc as a typical radius of supershell, and $v$ = 40 \kms ,
$g \sim 3.6 \times 10^{-8}$ cm s$^{-2}$. For full fragmentation,
we assume the largest disturbances to be the size of the tickness of the
shell ($\sim 10$ pc). For $\rho_2 \gg \rho_1$,
$\tau_{R-T} \sim 10^6$ yrs; the fragmentation time
is comparable to the sound crossing time of the
shell (Vishniac 1983). Thus
R-T instabilities develop in expanding shells on a timescale at least
ten times shorter than the lifetime of the parent OB associations.

A similar approach allows to derive the e-folding
growth-time of the Kelvin-Helmhotz instability waves (see also Fleck 1984).
When a fluid moves with differential velocity V with respect to another,
Kelvin-Helmhotz instability waves will develop. A range of unstable
wavelengths will grow, where (Chandrasekhar 1961)

\centerline {$ \lambda_{max} = 2 \pi \ V^2 \ (\rho_-/\rho_+)\ g^{-1}$}

\noindent where $\rho_-/\rho_+$ is the gas density ratio across the
boundary and $g$ is the acceleration at the boundary. The rise
time for instability of wavelength $\lambda$ is (Chandrasekhar 1961)

\centerline {$ \tau_{K-H} \ \sim \ {\lambda \over 2\pi} (\rho_+/\rho_-)^{1/2}\
V^{-1}.$}

\noindent To allow complete mixing of existing inhomogeneities throughout
a well-developped \hii region, we compute the time for the wavelength
of the largest disturbance to be or the order of the radius of a large
\hii region. Because of the square-root dependence on density, we
make the approximation  $n_+ \sim n_-$ within the nebular gas; for $\lambda
\sim$ 100 pc, and
V = 10  \kms , $\tau _{K-H} = 1.5 \times 10^6$ yrs, which again is much
smaller than the lifetime of a typical OB association.
With hot massive stars eroding the associated molecular cloud, jet generation
by radiation-driven
ionization fronts interacting with nonuniform neutral
cloud edges (Sanford, Whitaker \& Klein 1982) may add turbulent
energy to the vortices generated by R-T and K-H instabilities.
These considerations on instability growth show that abundance inhomogeneties
introduced by evolving massive star forming
regions should have the time to fully mix before the death of the
\hii region, even assuming that
most new elements might be ejected only at the end of the lifetime of
the stellar associations; evolutionary models of coeval OB associations
indicate that most of the oxygen is ejected at ages between 4 and 10 millions
years of age (Lequeux \etal 1981).

We summarize in Table 1 the various typical linear and time characteristic
scales
of processes responsible for mixing and erasing inhomogeneities in the
gaseous interstellar medium of disk galaxies.

\begin {center}
{\bf Table 1. Mixing Mechanisms in the Interstellar Medium}

\begin {tabular}{clrl} \hline
Scale & Mechanism  & timescale & Remarks \\
(pc) &  & (yr) \\
 \hline
$10^3 - 10^4$ & -- Turbulent transport& $< 10^9$ & Azimuthal\\
& in differential rotation & & homogenization \\ \\
& -- Bar induced & $10^9$ & Radial mixing \\
& radial flows & & \\ \\
& -- Triggered star formation &$ > 4 \times 10^9$& Radial and azimuthal \\
& from expanding supershell && homogenization (Sm, SBm)\\ \\
$10^2 - 10^3$& -- SNR and supershell in a& $ < 10^8$ & Intermediate \\
&differentially rotating disk& & scale mixing\\ \\
$1 - 10^3$ & {-- Turbulent diffusion$^*$}  & &\\
& $\bullet$H II region & $4 \times 10^7$ & Mixing of fresh nuclei \\
& $\bullet$H I cloud & $4 \times 10^8$ & Mixing within cold clouds \\
& $\bullet$H$_2$ cloud & $1.6 \times 10^9$ & Mixing of dormant clouds \\ \\
& -- R-T and K-H instabilities & $1.5 \times 10^6$ & Local mixing \\
& in star forming regions & & of fresh nuclei\\  \hline
\end {tabular}
\end {center}
\noindent (*) Timescale calculated for clouds of 1 kpc in diameter;
H$_2$ clouds are more likely of the order 100 pc.

\baselineskip 0.7cm
The main conclusions of our analysis are as follows: ($i$) Because of the
hierarchical continuity in the linear and time scales of mixing processes
($i.e.$ timescales of mechanisms mixing at small scales are much shorter
than the those operating at larger scales), one should expect a relatively
well mixed interstellar medium in disk galaxies, $i.e.$ one where
$\mid \delta$O/H$\mid$ $\ll \ 10^{-4}$ on scales $l \geq 1$ kpc. ($ii$)
We reckon that the discrepancy betweeen observed and expected abundance
fluctuations is significantly large, observed variations
being 5 - 10 times larger than expected ones ($iii$).
Our analysis of the relative
importance of large-scale and small-scale processes, $i.e.$ cloud motions
and collisions $vs.$ turbulent processes, leads to obvious differences
in mixing efficiency between dwarf and large galaxies. In particular, the
weakness
of the rotational velocity field in dwarf galaxies, the main agent for
mixing neutral \hi and molecular gas, and selective loss
through galactic winds can lead to large abundance
discontinuities in the smaller galaxies as already implied
in section 3.2.  Processes associated with massive
star formation are extremely efficient at mixing hot and warm ionized
gas; as a corollary, mixing is much less efficient in the cold neutral and
molecular gas, mixing of these latter phases being done by galactic
scale stirring.  As a consequence, large abundance spatial fluctuations are
more likely to be found in dwarf galaxies with long dormant phases
between star forming episodes.

\bigskip

\noindent
3.4 The large abundance discontinuity in I Zw 18 and the C:N:O problem

\bigskip

Because dwarf galaxies lack the stirring effect of large rotational
velocity fields and because they have gone through very few massive star
formation events, the main mixing processes associated with
vigorous star formation are not operating in
these systems.  The presence of substantial amount of \hi gas that is not
rotationally supported in the faint dwarf galaxies is an unsolved puzzle (Lo
\etal 1993). The case of the super-metal-poor low-mass galaxy I Zw 18 is
particularly
interesting to review.

Kunth and Sargent (1986) have proposed the view that giant \hii regions are
self-enriched. In such a case the new heavy elements ejecta originating from
stellar winds and supernovae of type II (SNII) initially mix exclusively
with the ionised gas
in the \hii zone, waiting for further mixing with the cold gas during the
long interburst phase. Morever as the authors note the closed-box model
leads to over-enrichment of oxygen and only one burst is enough to
produce the O/H in I Zw 18. Their suggestion has received a strong
support from the recent abundance determination in the \hi gas of I Zw 18
indicating that a previous burst could have occurred in the past leading
to enrich the \hi up to only 1/1000 the solar value after mixing
in a time scale of about 10$^9$ years ( Kunth et al. 1994), a timescale
long enough to allow turbulent diffusion (section 3.3.1) to homogenize
a cold cloud 1 kpc in diameter. Not only I Zw 18 is among the
lowest abundance objects, but it also displays the largest
abundance discontinuity.

While N/O is proportional to O/H for relatively massive
galaxies, low-mass (and oxygen poor) galaxies
have on average smaller N/O ratios than oxygen rich ones, but some oxygen
poor galaxies have the same N/O as oxygen rich ones; for 12 + log O/H between
7.5 and 8.5, log N/O is scattered between -1.0 and -2.0. Some of the N/O
scatter at low O/H is thought to be real, and  N/O is certainly well
measured in I Zw 18. Although Pagel $etal.$ (1992) and Skillman \& Kennicutt
(1993) find a lower value of N/O than
Dufour $etal.$ (1988) in I Zw 18, log N/O is certainly $not$ --2.0, --3.0
or less! To explain the observed scatter of N/O versus O/H for metal poor
galaxies, Pilyugin (1992) built up models in which mixing is controlled by
two processes:
self-enrichment and galactic winds due to SNII.
According to Pilyugin, each generation of stars contributes to the
chemical enrichment of the ISM and metals mix into the whole
galaxy. When starbursts begin, the N/O is unmodified, but as they
evolve, large amounts of O are produced giving lower N/O ratios and
an increase of O/H. On a short time scale, oxygen is enhanced more than the
nitrogen because the bulk of nitrogen is produced by stars that live
longer than those producing the bulk of oxygen. However the closed
box model alone cannot explain the general N/O vs O/H trend nor the
scatter.

A more complicated model must be assumed. Pilyugin introduces
galactic stellar winds  as proposed by Matteucci \& Chiosi (1983) and
Matteucci \& Tosi (1985) where
some oxygen-rich ejecta from SNII leave the galaxy before full mixing with the
gas
is completed (Russell \etal 1988; de Young and Gallagher 1990). By acounting
for galactic
stellar winds, better agreement with the scattered N/O vs O/H diagram is
obtained. The key features in this modified model are that oxygen-poor
galaxies are objects with efficient $enriched$ galactic winds, and that
they are in advanced stages of a star formation burst. Similar models
developped by Marconi \etal (1994) reproduce the position of I Zw 18 using
one single burst that started  $5 \times 10^7$ years ago coupled with a high
differential wind in which O is preferentially expelled compared to N.
Evidence for this type of winds may be found in the nearby small
starburst region NGC 2363 (Roy \etal 1992) where broad components in
the main nebular lines probably arise from a superwind with
velocity greater than 1000 km s$^{-1}$. The existence of a similar superwind
in I Zw 18 is suggested by Skillman \& Kennicutt (1993).  Consequently
the loss of metals through winds may be a key to understanding the chemical
evolution of dwarf galaxies.

On the other hand Pantelaki and Clayton (1987) suggest a different scenario.
Based on the high ratios of N/O and C/O in I Zw 18, they rule out the
possibility
that  the present starburst in I Zw 18 is
the first one; moreover they shunt aside the possibility that the \hii regions
are contaminated by supernova ejecta from this very present burst. Instead they
consider a situation in
which previous bursts account for a hot gaseous phase (T $\sim 10^6$ K)
surrounding
a \hi complex of small masses orbiting around the central stellar cores. These
\hi clouds eventually collide, giving rise to new starbursts (from
initial abundances close to primordial) whereas the hot gas could contain
large O and C concentrations; about one percent of this gas may have mixed with
the \hi. Each starburst spreads over $10^7$ years and SNII disperse material
into the hot
medium. Between bursts, intermediate mass stars produce SNI and a lot of
oxygen-free ejecta, thus O decreases in between bursts whereas C and N do not.
However, Lo \etal (1993) have pointed out that the maintainance of a large
volume filling factor of a hot
phase in the ISM depends on a high frequency of supernovae which is
unlikely for dwarf galaxies. Finally, from the spectroscopy of a new sample of
very
metal-poor galaxies (chosen from the SBS survey), Thuan \etal (1994) find a
straight N/O value, with very little dispersion, and independent of the O/H
value.  This could be a further indication
that C:N:O ratios in metal-poor galaxies undergoing their first bursts
are indicative of genuine primary elements (see  Maeder 1992 for a discussion
of the C yield
and Marconi \etal 1994 for the N prescription). In this case,
the single burst hypothesis and the youth of I Zw 18 would not remain a
problem.

\bigskip
\noindent {\bf 4. Discussion}

\bigskip

We have shown in the previous sections that small and medium-scale dynamical
processes in the ionized medium, combined with turbulent
transport in the shear flow of differential rotation should reduce
azimuthal inhomogeneities in disk galaxies, in a timescale of less than
$10^9$ yrs, to a level where they could
hardly be measured with present techniques. In low-mass galaxies with
little rotational shear, such mechanisms are much less
efficient than in large disk galaxies, or do not operate at all, and
large abundance discontinuities are likely to survive more than
$10^9$ years.

Assuming that the amplitude of observed O/H abundance fluctuations in
large disk galaxies and magellanic irregulars are significantly larger
than expected, we propose two complementary
effects that may act to build up, or maintain, large abundance fluctuations
in the ISM.  The first effect is one of retention of newly enriched
material in regions prone to undergo relatively quickly successive episodes
of star formation. Fluctuations can  be created
by localized bursts of star formation (whether or not triggering
is involved). Implicit to sequential star formation is some confinement  of
stellar ejecta:
newly produced elements are ``trapped'' in SN remnants and superbubbles, which
are the privileged sites for new star formation because
their higher densities make them prone to quicker collapse under
gravitational instability.
The second effect, as suggested by Edvardsson $etal.$ (1993), is infall
of relatively unprocessed gas, where the timescale of infall events is
shorter than the epicyclic mixing time described in section 3.1. Let us
examine each of these mechanisms.

\bigskip

\noindent 4.1 Abundance inhomogeneities from triggered star formation?

\bigskip

Franco (1992) has shown that shear due to differential rotation has
two important effects on expanding supershells: first, it distorts
the shape of the remnant as illustrated by the numerical
simulation of Palo\u us \etal (1990), and, second, it changes the
distribution of mass due to the epicyclic motion of the particles
in the expanding shell which creates particle flow toward the
tips of the distorted remnant, where the shell mass tends to
be accumulated; these preferentially enriched tips then become the preferred
locations for molecular cloud production and star-forming clouds.  The
consequence of triggered star formation is that some clouds spent much
less time in the dormant phase, thus newly ejected elements are retained for
the
to-be-borne stars.

The amplitude of O/H fluctuations will depend on the relative
importance of $stimulated$ star formation with respect to $spontaneous$
star formation
driven by gravitational instabilities. This requires the time scale
for gravitational instability to be much shorter for expanding gas
than for stationary gas. Following
Elmegreen (1994), this  implies $(G \rho_0 M^2)^{-1/2}$
$\ll$ $(G \rho_0)^{-1/2}$ for expanding rings, or
$(G \rho_0 M)^{-1/2}$ $\ll$ $(G \rho_0)^{-1/2}$
for expanding shells. We recall that $M = V/c$ is a dimensionless
quantity which measures the ring or shell compression or thickness; when
$M$ is large, the shell is thin and the collapse is rapid because of high
density.  Once formed the lifetime of an expanding shell (or ring) driven by
evolving
associations of massive stars is determined by the time for Coriolis force
due to galactic rotation and shear to distort and erase the blown cavity
(Edmunds 1975); this time is $\tau_{Coriolis}$ = 2.5/$\kappa$  (Palo\u us \etal
1990; Franco 1992), $\kappa$ being the epicyclic frequency. For the solar
neighborhood,
$\tau_{Coriolis}$ $\sim$ 10$^8$ yr. We showed is section 3.2 that
gravitational collapse of expanding shells $can$ take place in much less that
this time;
thus it is plausible to consider a sequence of star forming activities
involving a same given portion of the disk.
To explain large abundance fluctuations, one requires some portions
of a galaxy disk to obey (Elmegreen 1994)

\bigskip

\centerline {${1.5 \over (G \rho_0)^{1/2} M} <  {2.5 \over \kappa}$, \ \ \ for
rings}

or

\centerline {${1.25 \over (G \rho_0 M)^{1/2}} < {2.5 \over \kappa}$, \ \ \ for
shells.}

\bigskip

\noindent These relations must be obeyed for each episode of stimulated
star formation; a whole sequence of triggered of episodes of massive star
formation may last longer than $\tau_{Coriolis}$.

Furthermore, we suggest that the $ring$ relation of above is applicable to the
disks of
massive galaxies because most of the accumulated material originally
in the interior of the blown cavity will stay in the galactic plane; the
$shell$ relation on the other hand, is applicable to low-mass galaxies,
 where the restoring force is weaker due to the shallow
gravitational potential and material can move to the halo easily.  Thus for the
same parameter values, triggered star formation takes a longer time (by a
factor $M^{1/2}$) to occur in expanding shells
compared to rings. In addition, the compression of an
expanding shell would be weaker
in metal deficient objects like I Zw 18 because of less efficient
cooling; the shells in dwarf galaxies are then thick and
their corresponding values of $M$ small; this delays
their collapse further. Furthermore, dwarf galaxies are ``fat'' sytems
with a very clumpy \hi distribution (Lo \etal 1993); evolving superbubbles
from OB association will grow faster in directions with the lowest
column densities to the intergalactic medium. Thus for
superbubbles expanding in low-mass galaxies, most of the
internal pressure may be released through a chimney or a blowout (Ikeuchi
1987).
The blowout phenomenon is a transformation of
internal energy to kinetic energy, and the blowout power corresponds to
a reduced mechanical power in the plane of the galaxy (Schiano 1985).
The C:N:O abundance ratios in I Zw 18, as we have shown, are certainly
indicative of losses through some sort of wind.
To summarize,  triggered star formation is not  effective in very low mass
systems such as I Zw 18 because $(i)$ the time for gravitational collapse of an
expanding spherical shell
is longer than to collapse a ring, $(ii)$ low metallicity results in
lower compression of shells (small $M$), thus again in slower collapse,
and $(iii)$ the loss of material by galactic wind and of internal pressure
through blowout or chimney quenches the piston effect in the ISM of the galaxy.

In larger and more metal-rich disk galaxies, expansion as a ring, high
compression and low probabiblity of blowout make
 triggered star formation an effective process.
To estimate the amplitude of abundance fluctuations built up
through retention of enriched gas in sequential events of stimulated star
formation,
we assume that
nucleosynthetic products ejected in the ISM have been well-mixed by now,
$except$ for those injected in the last 1 - 2 $\times 10^9$ yrs.
We associate the region with the highest O/H with regions where
$both$ stimulated and spontaneous star formation have contributed,
while regions with the lowest O/H correspond to regions affected by
spontaneous star formation only. In our scenario, we do consider
that most star formation in disks is spontaneous on galactic scale and
lifetime, but that some pockets can undergo  vigorous
sequences of starbursts. If we do not allow the new nucleosynthetic
products to quickly mix with the larger-scale surrounding, the most extreme
discrepancies of abundances will be observed
between the portions enriched by both spontaneous and stimulated on one hand,
and the regions enriched only from spontaneous star formation on the other
hand.
The difference can be described approximately as

\bigskip

\centerline {${O/H_{max}\over O/H_{min}} =  [\ t_1 + \ t_2 \ (SF_{stim}/
                SF_{spon})]\ \tau_G^{-1}$,}

\bigskip

\noindent where $t_1$ is the duration of galactic enrichment period
whose products are fully mixed, and $t_2$ is the duration where mixing
has not had time yet to take place; $\tau_G$ is the Galaxy age (10$^{10}$ yrs).
 Evidently,
the shorter $t_2$ is, the higher the ratio $SF_{stim}/SF_{spon}$  must
be. For $t_2$ of $2 \times 10^9$, $10^9$ and $10^8$ yrs, the ratio is
5, 10 and 50 to explain variation by a factor of 2 in O/H this way.
We suggest that a period of $t_2$ = $10^9$ yrs is plausible; remember
that we consider the factor of 2 abundance discontinuity to be an extreme;
one can imagine a given group of clouds caught, during the last
$10^9$ yrs (four Galactic rotations) in a sequence
about 10 episodes of star formation while some other group, at an
equivalent galactocentric distance, would have undergone only one or two such
events. We exclude $t_2$
= $2 \times 10^9$ yrs because it is difficult for groups of clouds
to keep their identity over such a long duration (see section 3.1);
shorter periods such as $t_2$ = $10^8$ yrs are also excluded because of
the extreme rate of stimulated SF required.

\bigskip

\noindent 4.2 Infall and ``splatter'' as the source of inhomogeneities

Infall of low-metallicity material on the galactic disk is also an attractive
mechanism to account for the presence
of large abundance fluctuations in disk galaxies.  First infall solves
other problems related to galactic chemical evolution ($e.g.$
the G-dwarf problem). Second  it is a simple and straightforward
way to explain why the metal abundance of star clusters located
within 1 kpc of each other can differ by as much as a factor
of 5 in abundances (Rolleston \etal 1994), why the oxygen abundance in the
Orion Nebula is
only 1/2 solar while being about 4.5 billion years younger than our Sun,
and why there is so much observed scatter of [Fe/H] in nearby clusters.

Mayor \& Vigroux (1981) and Pitts \& Tayler (1989) have discussed
in detailed the effect of infall of matter on the dynamics and
chemistry of galaxy disks; mass accretion rates of $\leq 1\ M_\odot$/yr
are usually implied for the whole galaxy. Franco \etal (1988) have suggested
that
the Orion and Monoceros molecular cloud complexes result
from the interaction of high-velocity clouds and the disk of the Galaxy.
Although falling gas may be in large part recycled material participating in
the circulation
between the halo and the disk, there are evidences that some
high-velocity clouds may be of extragalactic origin (Mirabel 1989). Some
of these have low
abundances, as demonstrated by Kunth \etal (1994) who measured
O/H and Si/H to be about 1/10 the abundance of the local
interstellar gas in a Galactic high velocity cloud. When impacting
on the disk, the colliding remnants would give rise to molecular
cloud formation followed by the birth of massive stars (Franco \etal 1988).
The stars and the ionized gas resulting from such collisions would then
display anormal abundances compared to the expected abundances
from the normally inferred history of star formation at that
galactocentric distance; it is certainly easy to produce regions with
half the solar abundances like Orion.  If the Magellanic Stream is
a source of infall, the abundance in the LMC being about 1/3 solar, this
may also give rise to local abundance anomalies depending on where the
accreting gas
impacts on the Galaxy.

\bigskip
\noindent {\bf 5. Conclusion}

\bigskip

We have reviewed various stellar and interstellar abundance
indicators of metallicity of the interstellar medium of gas-rich
galaxies:  there is a growing body of observationnnal data showing
that there exist significant spatial abundance
fluctuations; the amplitude of the observed fluctuations appears 5 - 10
times larger than expected, given the apparent high efficiency of
mixing in the interstellar
medium of most galaxies.  Indeed, examining
the processes capable of moving, re-distributing and mixing the ISM,
we have shown that
turbulent transport in the shear flow of a differentially rotating disk
will efficiently mix the ISM of large disk galaxies so that azimuthal
inhomogeneities will persist less than $10^9$ years. On scales of 1 kpc or
less, mixing is achieved in good part by gas motions generated by star
formation.
The absence of abundance gradient
magellanic irregulars is probably due to the radial homogenizing action
of a bar.

To explain the apparent large amplitude of  spatial abundance
fluctuations in massive disk galaxies, we suggest retention of enriched ejecta
in
evolving structures favored by stimulated or triggered star formation, and
possibly infall of clouds with low abundance content of metals.
 We have also shown that the largest  fluctuations are expected in low mass
galaxies. This is clearly illustrated by
the largest abundance discontinuity found in the low-mass galaxy
I Zw 18; it is explained by the long dormancy between episodes of
star formation, due the inefficiency of triggered star
formation and the lack of powerful large-scale stirring mechanisms
like differential rotation.

\bigskip

\bigskip

We thank Mike G. Edmunds, Pierre Martin, Serge Pineault and Gilles
Joncas for helpful discussions. JRR thanks Chantal Balkowski and
the Observatoire de Paris-Meudon staff for their kind hospitality
while on research leave from Universit\'e Laval.
The research of JRR was funded by the Centre National de la Recherche
Scientifique of France, Universit\'e Laval, the Fonds pour la Formation
de Chercheurs et l'Aide \`a la Recherche of Quebec and
the Natural Sciences and Engineering Research Council of Canada.

\newpage
\large
\noindent $References$
\normalsize
\begin{description}

\item[] Allen, D. A., Burton, M. G., 1993, Nature, 363, 54
\item[] Arsenault, R., Roy, J.-R., 1988, A\&A, 201, 199
\item[] Baldwin, J. A., Ferland, G. J., Martin, P. G., Corbin, M. R., Cota, S.
A.,
Peterson, B. M., Slettebak, A., 1991, ApJ, 374, 580
\item[] Bateman, N. P. T., Larson, R. B., 1993, ApJ, 407, 634
\item[] Belley, J., Roy, J.-R., 1992, ApJS, 78, 61
\item[] Binney, J., \& Tremaine, S., 1987, Galactic Dynamics.
Princeton University Press, Pinceton
\item[] Boesgaard, A. M. 1989, ApJ, 336, 798
\item[] Boulanger, F., Viallefond, F., 1992, A\&A, 266, 37
\item[] Brinks, E., Bajaja, E., 1986, A\&A, 169, 14
\item[] Carlberg, R. G., Dawson, P. C., Hsu, T., Vandenberg, D. A., 1985,
ApJ, 294, 674
\item[] Castor, J., McCray, R., Weaver, R., 1975, ApJ, 200, L107
\item[] Chandrasekhar, S. 1961,  Hydrodynamic and Hydromagnetic
Stability. Oxford University Press, Oxford
\item[] Court\`es, G., Petit, H., Sivan, J.-P., Dodonov, S., Petit, M. 1987,
A\&A, 174, 28
\item[] Deul, E. R., Den Hartog, R. H., 1990, A\&A, 229, 362
\item[] de Young, D.S., Gallagher III, J.S., 1990, ApJ, 356, L15
\item[] Dopita, M., Mathewson, D. S., Ford, V. L., 1985, ApJ, 297, 599
\item[] Dufour, R. J. 1986, PASP, 98, 1025
\item[] Dufour, R. J., Harlow, W. V., 1977, ApJ, 216, 706
\item[] Dufour, R. J., Garnett, D. R., Shields, G. A., 1988, ApJ, 332, 752
\item[] Edmunds, M. G., 1975, ApSS, 32, 483
\item[] Edmunds, M. G., Roy, J.-R., 1993, MNRAS, 261, L17
\item[] Edvardsson, B., Andersen, J., Gustafsson, Lambert, D. L., Nissen,
P. E., Tomkin, J., 1993, A\&A, 275, 101
\item[] Elmegreen B. G. 1992, In: Tenorio-Tagle, G., Prieto, M., Sanchez, F.
(eds.) Star Formatiion in Stellar
Systems. Cambridge Univ. Press, Cambridge, p. 381
\item[] Elmegreen, B. G. 1994, ApJ May 20
\item[] Falgarone, E., Phillips, T. O., 1991, In: Falgarone, E. {\it et al.}
(eds.)
Proc. IAU Symp. 147,  Fragmentation of Molecular Clouds and Star Formation
Reidel, Dordrecht, p. 119
\item[] Fich, M., Silkey, M., 1991, ApJ, 366, 107
\item[] Fitzsimmons, A., Brown, P. J. F., Dufton, P. L., Lennon, D. J.,
1990, A\&A, 232, 437
\item[] Fleck, R. C., 1984, AJ, 89, 506
\item[] Franco, J., 1992, In: Tenorio-Tagle, G., Prieto, M., and Sanchez, F.
(eds.)
Star Formation in Stellar
Systems, Cambridge Univ. Press, Cambridge, p. 515
\item[] Franco, J., Tenorio-Tagle, G., Bodenheimer, P., R\'ozyczka,
Mirabel, F., 1988, ApJ, 333, 826
\item[] Fran\c cois, P., Matteucci, F., 1993, A\&A, 280, 136
\item[] Friedli, D., Benz, W., 1993, A\&A, 268, 65
\item[] Friedli, D., Benz, W., Kennicutt, R., 1994, ApJL, submitted
\item[] Gehren, T., Nissen, P. E., Kudritzki, R. P., Butler, K., 1985,
In: Proc. ESO Workshop on Production and Distribution of CNO Elements,
ESO, Garching, p. 171
\item[] Gilmore, G., 1989, In: Buser, R.,
King, I. R. (eds.) The Milky Way as a Galaxy. Univ. Science Books, Mill Valley
Ca., p. 281
\item[] Green, D. A., 1984, MNRAS, 209, 449
\item[] Hausman, M. A., 1981, ApJ, 245, 72
\item[]Igumentshchev, I. V., Shustov, B. M., Tutukov, A. V., 1990,
A\&A, 234, 396
\item[] Ikeuchi, S., 1987, In: Thuan, T. X., Montmerle, T., Tran Thanh Van, J.
(eds.) Starbursts and Galaxy Evolution,  Editions Fronti\`eres, Gif-sur-Yvette,
p. 27
\item[] Kennicutt, R. C., 1992, In:  Tenorio-Tagle, G., Prieto, M., Sanche, F.
(eds.) Star Formation in Stellar Systems, Cambridge Univ. Press, Cambridge, p.
191
\item[] Kulkarni, S. R., Heiles, C. 1988, In: Verschuur, G. L.,  K. I.
Kellerman, (eds.)  Galactic and Extragalactic Radioastronomy.
Springer-Verlag, Berlin, p. 95
\item[] Kunth, D., Sargent, W. L. W., 1986, ApJ, 300, 496
\item[]Kunth, D., Lequeux, J., Sargent, W. L. W., Viallefond, F.,
 1994, A\&A, 282, 709
\item[]Lacey, C. G., Fall, S. M., 1985, ApJ, 240, 154
\item[] Lennon, D. J., Dufton, P. L., Fitzsimmons, A., Gehren, T., Nissen,
P. E., 1990, A\&A, 240, 349
\item[] Lequeux, J., Maucherat-Joubert, M., Deharveng, J. M.,  Kunth, D., 1981,
A\&A, 103, 305
\item[] Lo, K. Y., Sargent, W. L. W., Young, K., 1993, AJ, 106, 507
\item[] Maeder, A., 1992, A\&A, 264, 105
\item[] Malinie, G., Hartmann, D. H., Clayton, D. D., Mathews, G. J., 1993,
ApJ, 413, 633
\item[] Marconi, G., Matteucci, F., Tosi, M., 1994, MNRAS in press
\item[] Martimbeau, N., Carignan, C., Roy, J.-R., 1994, AJ, 107, 543
\item[] Martin, P., Roy, J.-R., 1994, ApJ, 424, 599
\item[] Matteucci, F., Chiosi, C., 1983 A\&A, 123, 121
\item[] Matteucci, F., Tosi, M., 1985, MNRAS, 217, 391
\item[] Mayor, M., Vigroux, L., 1981, AA, 98, 1
\item[] McCray, R. Kafatos, M., 1987, ApJ, 317, 190
\item[] Meaburn, J., Solomos, N., Laspias, V., Goudis, C., 1989,
A\&A, 225, 497
\item[] Mihalas, D., Binney, J., 1981, Galactic Astronomy, W. H. Freeman and
Co.,
San Francisco
\item[] Mirabel, I. F., 1989, In: Tenorio-Tagle, G., Moles, M., Melnick, J.
(eds.) Structure and Dynamics of the
Interstellar Medium. Springer-Verlag. Berlin, p. 396
\item[] Pagel, B. E. J., 1993,  eds. Baschek, B., Klare, G., Lequeux, J. (eds.)
 New Aspects of Magellanic Cloud
Research. Springer-Verlag, Berlin, p. 330
\item[] Pagel, B. E. J., Edmunds, M. G., Fosbury, R. A. E., Webster, B. L.,
1978,
MNRAS, 184, 569
\item[] Pagel, B. E. J., Simonson, E. A., Terlevich, R. J., Edmunds,
M. G., 1992, MNRAS, 255, 325
\item[] Palo\u us, J., Franco, J., Tenorio-Tagle, G., 1990, A\&A, 227, 175
\item[] Pantelaki, I., Clayton, D., 1987, In:  Thuan, T. X., Montmerle, T.,
Tran Thanh Van, J. (eds.) Starbursts and galaxy evolution.
Editions Frontieres, Gif-sur-Yvette, p.145
\item[]  Pilyugin, L.S., 1992, A\&A, 260, 58
\item[] Pitts, E., Tayler, R. J., 1989, MNRAS, 240, 373
\item[] Puche, D., Westpfahl, D., Brinks, E., Roy, J.-R., 1992, AJ, 103, 1841

\item[] Roberts, W. M. \& Hausman, M. A., 1984, ApJ, 277, 744
\item[] Rolleston, W. R. J., Dufton, P. L., Fitzsimmons, A., 1994, A\&A, 284,
72
\item[] Roy, J.-R., Boulesteix, J., Joncas, G., Grundseth, B., 1991,
ApJ, 367, 141
\item[] Roy, J.-R., Aub\'e, M., McCall, M. L., Dufour, R. J., 1992, ApJ,
386, 498
\item[] Russell S.C., Bessell M.S., Dopita, M.A.,  1988, In: Cayrel
de Strobel, G., Spite, M. (eds.) IAU Symp. 132, The Impact of
Very High S/N Spectroscopy on Stellar Physics. Reidel, Dordrecht, p. 545
\item[] Sandford, M. J., II, Whitaker, R. W., Klein, R. I.,
1982, ApJ, 260, 183
\item[] Schiano, A. V. R., 1985, ApJ, 299, 24
\item[] Schwartz, R. D. 1983, ARAA, 21, 209
\item[] Sellwood, J. A., Wilkinson, A., 1993, Reports on Progress in
Physics, 56, 173
\item[] Shaver, P. A., McGee, R. X., Newton, L. M., Danks, A. C., Pottasch,
S. R., 1983, MNRAS, 204, 53
\item[] Shore, S. N., 1992, An Introduction to Astrophysical Hydrodynamics,
Academic Press, Inc.,  San Diego
\item[] Skillman, E. D., Kennicutt, R. C., 1993, ApJ, 411, 655
\item[] Spitzer, L., 1978, Physical Processes in the Interstellar Medium,
Wiley-Interscience, New York
\item[] Struck-Marcell, C., 1991, ApJ, 368, 348
\item[] Tayler, R. J., 1993, Galaxies: Structure and Evolution,
Cambridge University Press, Cambridge
\item[] Tennekes, H. \& Lumley, J. L., 1983, A First Course in Turbulence,
The MIT Press, Cambridge, Mass.
\item[] Tenorio-Tagle, G., 1979, A\&A, 71, 59
\item[] Tenorio-Tagle, G., Bodenheimer, P., 1988, ARAA, 26, 145
\item[] Thuan, T. X., Izotov, Y. I., Lipovetsky, V. A., Pustilnik, S. A.,
1994, In:  ESO/OHP Workshop on Dwar Galaxies, in press
\item[] Vila-Costas, M. B., Edmunds, M. G., 1992, MNRAS, 259, 121
\item[] Vishniac, E. T., 1983, ApJ, 274, 125
\item[] Walsh, J. R., Roy, J.-R., 1989, ApJ, 341, 722
\item[] Zaritsky, D., Kennicutt, R. C., Huchra, J. P., 1994, ApJ, 420, 87
\end{description}
\end{document}